\documentclass[onecolumn,extra]{article}

\usepackage[dvips]{epsfig}
\usepackage{amssymb}
\usepackage{amsmath}
\usepackage{times}
\usepackage{color}

\usepackage[T1]{fontenc}

\newtheorem{theorem}{Theorem}[section]

\newcommand{\qed}{\nobreak \ifvmode \relax \else
      \ifdim\lastskip<1.5em \hskip-\lastskip
      \hskip1.5em plus0em minus0.5em \fi \nobreak
      \vrule height0.75em width0.5em depth0.25em\fi}

\begin{document}
\title
{Proof of validity of first-order travel estimates}
\author
{
Len Bos\\
Dipartimento di Informatica, Universit\`a di Verona, Italy\\
Michael A. Slawinski\\
Department of Earth Sciences, Memorial University of Newfoundand, St. John's, Newfoundland, Canada}
\maketitle

\begin{abstract}
In the seminal paper by Dahlen et al \cite{BD1} the authors formulate an important expression as a first-order estimate of traveltime delay. The authors left out a term which would at first glance seem nontrivial, on the basis that their intention was to derive the Fr\'{e}chet derivative linking the observed delay to the model perturbation (Nolet 2009, pers. comm.). Here we show that the derivation by \cite{BD1} results in a first-order estimate even without anticipating a Fr\'{e}chet derivative, but instead remaining deductively in their Taylor-series formulation. Although a mathematical technicality, this strengthens the result of \cite{BD1} by showing that it is intrinsically valid, requiring no external justification. We show also that ignoring the aforementioned term is not valid in general and needs to be supported by careful argument. 
\end{abstract}

\section{Introduction}
The purpose of this paper is to clarify and strengthen the 
derivation of formula (65) of the seminal paper by \cite{BD1}, in which the use of crosscorrelation to estimate traveltime delays is discussed. Formula (65) is a first-order estimate of $\delta \tau,$ the shift in the peak position of the crosscorrelation of observed and synthetic pulses. At a certain point in the derivation, which we explain in detail below, the Taylor series used by \cite{BD1} is of one order less than would be normally expected, and for a general equation it would not be possible to make the claim that the estimate is indeed of first order. However, we show that in this special context the claim is mathematically true for rather subtle reasons, without the need of justifying it by a subsequent geophysical application.
Even though, certain subtleties might be of no interest to geophysicists using the results of \cite{BD1}, we offer this proof in the context of geomathematics and in a spirit of respect for the contributions of the late Tony Dahlen.

Let us describe the derivation in question. The main entity studied is the crosscorrelation
\begin{equation}\label{58}
\Gamma(\tau)=\int\limits _{t_1}^{t_2} s(t-\tau)s^{\rm obs}(t)\mathrm{d}t,
\end{equation}
where $s$ is the synthetic signal and $s^{\rm obs}$ is the observed signal; this is equation (58) of \cite{BD1}. Subsequently, they assume that $s^{\rm obs}$ can be written as
\[s^{\rm obs}(t)=s(t)+\delta s(t),\]
where $\delta s(t)$ is a small perturbation; this is equation (59).
This permits them to write equation (60):
\begin{equation}\label{60}
\Gamma(\tau)=\gamma(\tau)+\delta\gamma(\tau),
\end{equation}
where 
\[\gamma(\tau)=\int\limits _{t_1}^{t_2} s(t-\tau)s(t)\mathrm{d}t\]
is the crosscorrelation of $s$ with itself, and
\begin{equation}\label{deltagamma}
\delta\gamma(\tau)=\int\limits _{t_1}^{t_2} s(t-\tau)\delta s(t)\mathrm{d}t
\end{equation}
is the crosscorrelation of the synthetic signal and the perturbation.

The maximum of the autocorrelation, $\gamma(\tau),$ is at $\tau=0$; the goal of  \cite{BD1} is to give a first-order estimate of the delay
given by the maximum of $\Gamma(\tau),$ which they call $\delta \tau.$ They obtain this estimate by taking the Taylor series  of $\Gamma(\tau)$ about $\tau=0$, to the second order, differentiating this second-order series, and taking $\delta \tau$ to be given by the
zero of this derivative. Herein is a technical subtlety.
The second-order Taylor series should be 

\begin{equation}\label{trueTaylor}
\Gamma(\tau)=\gamma(0)+\partial_\tau\gamma(0)\tau+\frac{1}{2}\partial_{\tau\tau}\gamma(0)\tau^2+\delta\gamma(0)+\partial_{\tau}\delta\gamma(0)\tau+\frac{1}{2}\partial_{\tau\tau}\delta\gamma(0)\tau^2 + \cdots,
\end{equation}
so  that 
\[
\partial_\tau\Gamma(\tau)= \partial_\tau\gamma(0)+\partial_{\tau\tau}\gamma(0)\tau+\partial_{\tau}\delta\gamma(0)+\partial_{\tau\tau}\delta\gamma(0)\tau + \cdots,
\]
and the critical point of $\Gamma(\tau),$ using the fact that
$\partial_\tau\gamma(0)=0,$ would be
\begin{equation}\label{truedeltatau}
\delta\tau\approx -{\partial_\tau \delta\gamma(0)\over \partial_{\tau\tau}\gamma(0)+\partial_{\tau\tau}\delta\gamma(0)}. 
\end{equation}
However, instead of the second-order series (\ref{trueTaylor}), \cite{BD1} use
\begin{equation}\label{usedTaylor}
\Gamma(\tau)=\gamma(0)+\partial_\tau\gamma(0)\tau+\frac{1}{2}\partial_{\tau\tau}\gamma(0)\tau^2+\delta\gamma(0)+\partial_{\tau}\delta\gamma(0)\tau + \cdots,
\end{equation}
which is their equation (63); in other words, they neglect
\begin{equation}\label{missingterm}
\frac{1}{2}\partial_{\tau\tau}\delta\gamma(0)\tau^2,
\end{equation}
which results in 
\begin{equation}\label{65}
\delta\tau\approx -{\partial_\tau \delta\gamma(0)\over \partial_{\tau\tau}\gamma(0)},
\end{equation}
which is estimate (65) of \cite{BD1}.

The absence of the term (\ref{missingterm}) is not trivial mathematically, even if one might argue for it heuristically in the context of geophysical approximations. We have examined the documents of the late Tony Dahlen, which he sketched in preparation for \cite{BD1}, and which are available at Princeton University. Not having found a mathematical justification therein, we give one
in Sections \ref{firstorder} and \ref{conclusion}, below. Before doing so, however, we give an example that shows that care must be taken when using such a method to estimate critical points.

\section{Illustrative Example}
We give an example showing that small perturbations may cause critical points
to move great distances.
Consider
\[f(t)={1+at^2\over 1+t^2},\]
where
\[a:=1-{\epsilon^4\over(1+\epsilon^2)^2},\]
with $1>\epsilon>0$, which implies that $0<a<1$.
Thus, we can write
\[f(t)={(1+t^2)+(a-1)t^2\over 1+t^2}=1-(1-a){t^2\over 1+t^2},\]
and conclude that
\[a\le f(t)\le 1,\quad t\in\mathbb{R},\]
and that $f(t)$ has a unique maximum at $t=0$: $f(0)=1.$
Furthermore,
\[f'(t)=2(a-1){t\over (1+t^2)^2},\]
which attains its maximum absolute value at $t=\pm \sqrt{3}/3$:
\[\left|f'(\pm\sqrt{3}/3)\right|={3\sqrt{3}\over 8}(1-a) = {3\sqrt{3}\over8}{\epsilon^4\over (1+\epsilon^2)^2};\]
in other words,
\begin{equation}\label{boundf'}
|f'(t)|\le {3\sqrt{3}\over8}{\epsilon^4\over (1+\epsilon^2)^2},\quad t\in\mathbb{R}.
\end{equation}

Now let
\[g(t)={\epsilon\over 1+(t-b)^2}\]
where $b:=1/\epsilon$, be a perturbation to $f(t)$.
Since 
\[|g(t)|\le \epsilon,\quad t\in\mathbb{R},\]
it is a small perturbation. 
Also,
\[g'(t)=2\epsilon {b-t\over (1+(t-b)^2)^2},\]
and we see that $g'(t)>0$ for $t\le 0.$ Hence, $h(t):=f(t)+g(t)$ is such that
\[h'(t)>0 \,\,\hbox{for}\,\, t<0,\]
and the perturbed function, $h$, has no negative critical points.

Since for $0\le t\le b/2,$ 
\begin{equation*}
|g'(t)|\ge {\epsilon b\over (1+b^2)^2}={1\over (1+1/\epsilon^2)^2}={\epsilon^4\over (1+\epsilon^2)^2}=1-a,
\end{equation*}
it follows that
\[|g'(t)|\ge 1-a>{3\sqrt{3}\over8}(1-a)\ge |f'(t)|,\]
and, consequently,
\[h'(t)\neq0,\qquad 0\le t\le b/2.\]
In other words, any critical points of the perturbed function,
$h(t)$, must be greater than $b/2=1/(2\epsilon),$ which can be arbitrarily large and distant from the original critical point of $f(t)$, which is at $t=0.$

\section{First-order estimates}
\label{firstorder}
As illustrated in the above example, it is important to understand what is happening if one estimates a critical point by using Taylor series in this way. A critical point, $c,$ of  $f(t)$ is a
root of its derivative, $f'(t).$ In the expectation that this critical point is
near $t=0$, it is reasonable to estimate $f(t)$ by its Taylor series about
$t=0,$
\[f(t)= f(0)+f'(0)t+{1\over2}f''(0)t^2+\cdots.\]
Differentiating, we get
\begin{equation}
f'(t)= f'(0)+f''(0)t+\cdots,\
\label{fprimet}
\end{equation}
and the estimate of the critical point is obtained by setting expression (\ref{fprimet}) to zero and solving for $t$ to get
\[c = -{f'(0)\over f''(0)},\]
which is analogous to expression (65) of \cite{BD1}.
Notice that this is one iteration, $c=t_1,$ of Newton's method for finding a root
of $f'(t)$ with $t_0=0$ as starting point. Hence the relevant error estimates 
are those for Newton's method.

\begin{figure*}
\centering
\includegraphics[height=8cm]{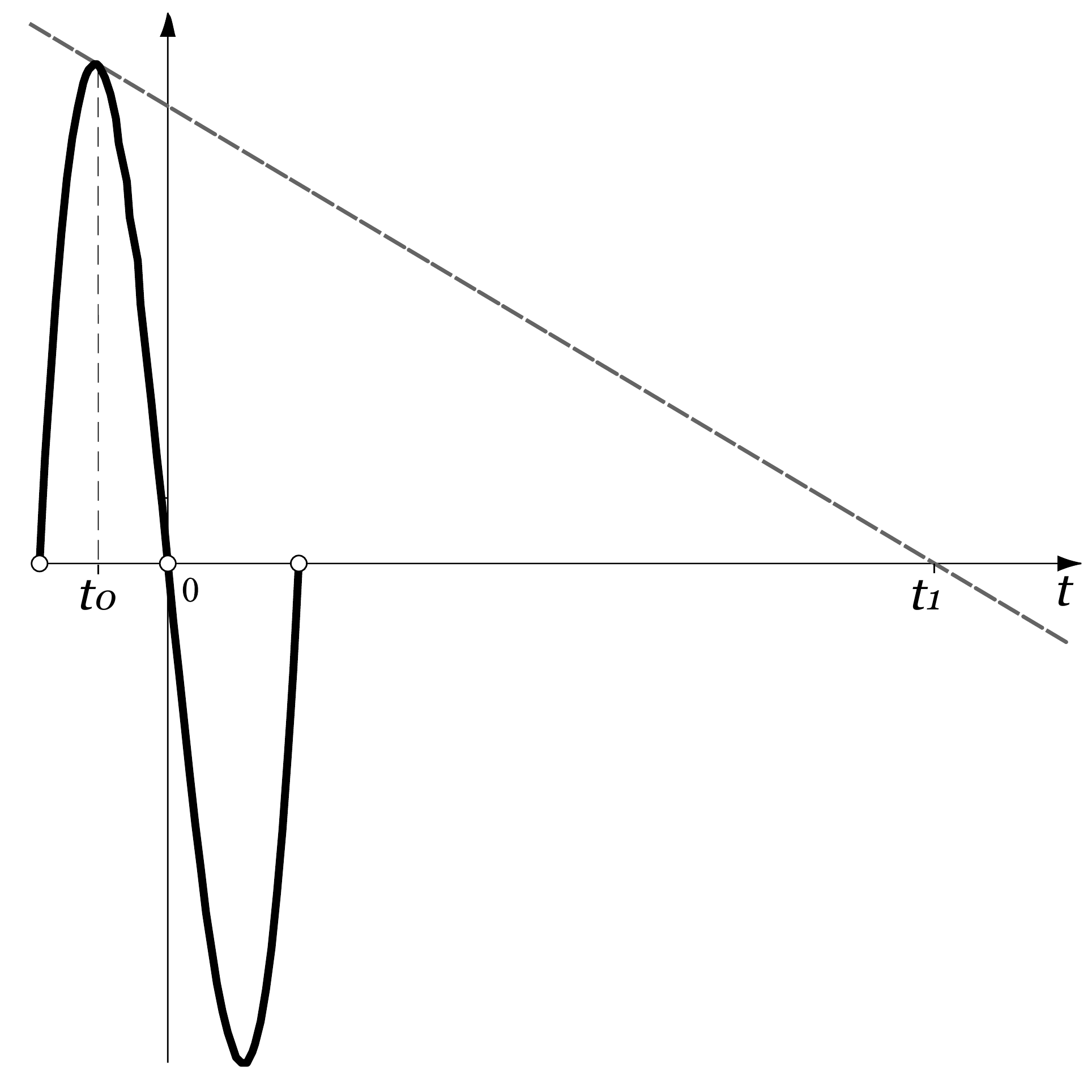}
\caption{An example of the Newton estimate being far from a root}
\label{fig1}
\end{figure*}

\begin{theorem} (\cite{ostr}, p. 56) Suppose that Newton's method is applied to the function $F(t)$ with starting point $t_0$ to obtain the estimate $t_1$ of the true root $t=z.$ Then
\[t_1-z=-{F''(\eta)\over 2(F'(\eta))^3}(F(t_0))^2,\]
for some $\eta$ between $z$ and $t_0.$
\end{theorem}
If $|t_0-z|$ is ``small'' then we may refine this error bound as follows. In this case $\eta\approx z$ so that 
\[-{F''(\eta)\over 2(F'(\eta))^3}\approx -{F''(z)\over 2(F'(z))^3},\]
which we may regard as approximately constant, provided $F'(z)\neq0.$ Furthermore,
we may expand
\[F(t_0)=F(z)+F'(z)(t_0-z)+\cdots=0+F'(z)(t_0-z)+\cdots\]
so that
\[|F(t_0)|\approx |F'(z)|\,|t_0-z|.\]
Hence we arrive at, under suitable technical assumptions, that
\begin{equation}\label{NewtonError}
|t_1-z|\le C |t_0-z|^2
\end{equation}
where $C$ is a constant that depends on $F,$ $z$ and $t_0.$ Notice that this is a restatement of the fact that Newton's method is quadratically convergent. But we must be careful; estimate
(\ref{NewtonError}) is local in nature --- it depends strongly on the
assumptions that $t_0$ is already close to $z$ and that $F'(z)$ is not close to $0.$  If this is not the case,
then there is nothing reasonable to say about $|t_1-z|.$ Figure \ref{fig1} illustrates this problem.

Returning to our problem, we use $F(\tau):=\partial_\tau\{\gamma(\tau)+\delta\gamma(\tau)\}$ with $\tau_0=0$, so that
\[\tau_1= -{\partial_\tau \delta\gamma(0)\over \partial_{\tau\tau}\gamma(0)+\partial_{\tau\tau}\delta\gamma(0)};\]
$\tau_1\equiv\delta\tau$ of expression (\ref{truedeltatau}). Hence we conclude
that indeed estimate (\ref{truedeltatau}) does provide a first-order approximation
to the delay, provided that the true critical point is close to
zero and that 
$\partial_{\tau\tau}\gamma(0)+\partial_{\tau\tau}\delta\gamma(0)$
is not close to zero. 

\section{Conclusion: Why estimate (\ref{65}) is of first order}
\label{conclusion}

We consider the difference between expression (\ref{truedeltatau}), which we know to be first-order, and estimate (\ref{65}) of \cite{BD1}, which we wish to show is also first-order. In fact we claim that the difference
between these two expressions is of {\it second} order, from which our claim follows. Specifically, we have
\begin{align}
\left|{\partial_\tau \delta\gamma(0)\over \partial_{\tau\tau}\gamma(0)+\partial_{\tau\tau}\delta\gamma(0)}-
{\partial_\tau \delta\gamma(0)\over \partial_{\tau\tau}\gamma(0)}\right|&=
|\partial_\tau\delta\gamma(0)|\,\left|
{1\over \partial_{\tau\tau}\gamma(0)+\partial_{\tau\tau}\delta\gamma(0)}
-{1\over \partial_{\tau\tau}\gamma(0)}\right|\nonumber\\
&=|\partial_\tau\delta\gamma(0)|\,\left|{\partial_{\tau\tau}\delta\gamma(0)
\over\partial_{\tau\tau}\gamma(0)(\partial_{\tau\tau}\gamma(0)
+\partial_{\tau\tau}\delta\gamma(0))}\right|.\label{diff1}
\end{align}
Following expression (\ref{deltagamma}), we write
\begin{equation*}
|\delta\gamma(0)|=\left|\int\limits _{t_1}^{t_2} s(t)\delta s(t)\mathrm{d}t\right|\le \max_{t_1\le t\le t_2}|\delta s(t)| \,\int\limits _{t_1}^{t_2} |s(t)|\mathrm{d}t,
\end{equation*}
which we rewrite symbolically as
\[|\delta\gamma(0)|\le C_1 |\delta s|.\]
Herein and below, $C_1,C_2$, etc. stand for generic constants.

Taking the derivative, we have
\begin{equation*}
|\partial_\tau\delta\gamma(0)|=\left|\int\limits _{t_1}^{t_2} \partial_ts(t)\delta s(t)\mathrm{d}t\right|\le \max_{t_1\le t\le t_2}|\delta s(t)| \,\int\limits _{t_1}^{t_2} |\partial_t s(t)|\mathrm{d}t,
\end{equation*}
which we rewrite symbolically as
\begin{equation}\label{deltasbnd1}
|\partial_\tau\delta\gamma(0)|\le C_2|\delta s|.
\end{equation}
Differentiating again, we have
\begin{equation}\label{deltasbnd2}
|\partial_{\tau\tau}\delta\gamma(0)|\le C_3|\delta s|.
\end{equation}

Using estimates (\ref{deltasbnd1}) and (\ref{deltasbnd2}) in expression
(\ref{diff1}), we have, for sufficiently small $|\delta s|$ and $ \partial_{\tau\tau}\gamma(0)\neq0,$
\[\left|{\partial_\tau \delta\gamma(0)\over \partial_{\tau\tau}\gamma(0)+\partial_{\tau\tau}\delta\gamma(0)}-
{\partial_\tau \delta\gamma(0)\over \partial_{\tau\tau}\gamma(0)}\right|\le C_4|\delta s|^2,\]
which means that the difference between the first-order
estimate (\ref{truedeltatau}) and estimate (\ref{65}) of \cite{BD1} is second-order in $\delta s$. Hence, we conclude that estimate (\ref{65}) is indeed a first-order estimate.

To demonstrate that estimate (\ref{65}) is of the first order, we remained in the context of steps (58) to (65) of \cite{BD1}, which are expressions (\ref{58}) and  (\ref{65}), herein. There is no need to anticipate Fr\'{e}chet derivatives that are to result from this derivation; in other words, there is no need to let $\delta\gamma$ tend to zero, as suggested by Nolet (2009, pers. comm.). Hence, the formulation herein, albeit technical, strengthens the first-order validity of estimate (\ref{65}) of \cite{BD1}.

\section*{Acknowledgments}
The authors are grateful for discussions with Qinya Liu, Tarje Nissen-Meyer, Daniel Peter, Michael Rochester and Jeroen Tromp, for email comments of Adam Baig, Shu-Huei Hung and Guust Nolet, for editorial help of David Dalton, for help of Frederik Simons in making Tony Dahlen's notes available. MS's research was supported by the Natural Sciences and Engineering Research Council of Canada, and a part of his work was done at Princeton University during a sabbatical year. The authors are grateful for the encouraging atmosphere at Casaleggio, where the ideas presented came to fruition.

\end{document}